\documentstyle[aps, epsf, psfig]{revtex}

\newcommand{\pa}{\partial}

\begin{document}

\draft
\title{On the Mixed Propagator Approach to $\rho -\omega$
Mixing
\thanks{This work is partly supported by the grant No.90103002 of National Natural Science
Foundation of China.}}
\author{Mu-Lin Yan\footnote{mlyan@ustc.edu.cn}, Ji-Hao Jiang}
\address{ CCST(World Lab), P.O.Box 8730, Beijing,100080, P.R.China\\
        and\\
 Center for Fundamental Physics,University of Science and Technology
 of China\thanks{Mailing address}  \\
  Hefei, Anhui 230026, China}
\author{Xiao-Jun Wang}
\address{Institution of Theoretical Physics, Beijing, 100080, P.R.China}
\maketitle

\begin{abstract}
Mixed Propagator (MP) approach to $\rho-\omega$ mixing is
discussed. It is found that under the pole-approximation
assumption the results of MP approach is not compatible both with
the effective Lagrangian theory and with the experiment
measurement criterion. To overcome these inconsistent, we propose
a new MP approach in which the physical states of $\rho$ and
$\omega$ are determined by the requirement of experimental
measurement to meson resonance. In terms of this new MP approach,
the EM pion form factor $F_\pi$ and form factors of $\rho^0
\rightarrow \pi^0 \gamma $ and of $\omega \rightarrow \pi^0
\gamma $ are derived. The results of $F_\pi$ are in good
agreement with data. The form factor of $\rho^0 \rightarrow \pi^0
\gamma $ exhibits a hidden charge-asymmetry enhancement effect
which agree with the prediction of the effective Lagrangian
theory.

\end{abstract}

\section*{}

$\rho^0-\omega$ mixing (or $\rho^0-\omega$ interference, or
interference between overlapping resonances, or charge asymmetry
caused by $\rho-\omega$ mixing and so on) has attracted much
interest during past four decades. Since 1961\cite{glashow},
there is a great deal of works on this subject in the literature
(see, for example, review articles\cite{oconnell,renard} and the
references within). Recently, we used model-independent effective
Lagrangian (eff-L) approach to discuss $\rho-$ and $\omega-$
anomaly-like processes of $\rho^0\rightarrow\pi^0\gamma$, and of
$\omega\rightarrow\pi^0\gamma$\cite{wjy}. It has been found that
there exist a hidden charge-asymmetry enhancement effect to
$\rho^0\rightarrow\pi^0\gamma$ due to $\rho^0-\omega$ mixing.
Namely, we have (by using the notations in ref\cite{wjy})
\begin{eqnarray}\label{1}
f_{\rho^0\pi^0\gamma}|_{\rm eff-L}&=&f^{(0)}_{\rho^0\pi^0\gamma}(1
  +\frac{3\Pi_{\rho\omega}(m_\rho^2)}
   {m_\rho^2-m_\omega^2+im_\omega\Gamma_\omega}),\nonumber\\
f_{\omega\pi^0\gamma}|_{\rm eff-L}&=&f^{(0)}_{\omega\pi^0\gamma}(1
  +{1\over 3}\frac{\Pi_{\rho\omega}(m_\omega^2)}
   {m_\omega^2-m_\rho^2+im_\rho\Gamma_\rho}),
\end{eqnarray}
where the subscript 'eff-L'  means the results obtained by in the
effective Lagrangian approach. Due to $ \Gamma_{\omega} <<
\Gamma_{\rho} $, we have
\begin{eqnarray}\label{101}
|f_{\rho^0\pi^0\gamma}|_{\rm eff-L}&>>&
f^{(0)}_{\rho^0\pi^0\gamma}=f_{\rho^\pm\pi^\pm\gamma}\\
|f_{\omega\pi^0\gamma}|_{\rm eff-L}&\sim
&f^{(0)}_{\omega\pi^0\gamma}.
\end{eqnarray}
  This
reflects an unusual charge asymmetry enhancement effect for $ \rho
\rightarrow \pi \gamma $. As addressed in\cite{wjy}, this hidden
effect has already been implied by experiment\cite{benayoun}.

On the other hand, There is a well-known quantum mechanics method
to deal with the $\rho-\omega$ mixing problem: the approach of
mixed propagator with pole approximation (shortly, we will call
it as pole-Mixed-Propagator approach, or pole-MP
approach)\cite{oconnell,renard}. This approach was developed even
before discovery of QCD. A question arisen here what are results
for $\rho^0 \rightarrow \pi^0 \gamma$ and $\omega \rightarrow
\pi^0 \gamma$ in pole-MP approach? The vector meson propagator is
given by (Renard representation)
\begin{eqnarray}\label{2}
D_{\mu\nu}(q^2)&=&\int
d^4xe^{-iq.x}\langle 0|T\{V_\mu(x)V_\nu(0)\}|0\rangle \nonumber \\
   &=&D(q^2)g_{\mu\nu}+{1 \over q^2}(D(0)-D(s))q_\mu q_\nu
\end{eqnarray}
where $s\equiv q^2$, and the propagator function $D(s)$ is
written in the following way
\begin{equation}\label{3}
D(s)={1\over s-W(s)}.
\end{equation}
For multi-vector-meson channels, $W(s)$ is complex mass-square
matrix with non-zero off-diagonal elements in general. The basic
assumption for pole-MP approach is that the physical states are
eigenvectors of $W(s)$, or the complex mass-square of physical
state is determined by the following equation
\begin{equation}\label{4}
det[s-W(s)]=0.
\end{equation}
For $\rho-\omega$ case, the solutions of this physical condition
equation are follows\cite{renard}
\begin{eqnarray}\label{5}
|\rho_p^0\rangle&=&|\rho_I^0\rangle +\epsilon |\omega_I\rangle,
\;\;\langle \tilde{\rho}^0_p|=\langle \rho_I^0|-\epsilon \langle \omega_I|
 \nonumber \\
|\omega_p\rangle&=&|\omega_I\rangle -\epsilon |\rho_I^0\rangle,
\;\;\langle \tilde{\omega}_p|=\langle \omega_I^0|+\epsilon
\langle \rho^0_I|
\end{eqnarray}
with
\begin{equation}\label{6}
\epsilon={\Pi_{\rho\omega}\over
m_\omega^2-m_\rho^2+i(m_\omega\Gamma_\omega -m_\rho\Gamma_\rho)},
\end{equation}
where $|\rho_p^0\rangle$ and $|\omega_p\rangle$ denote physical
$\rho^0-$ and $\omega-$ states under the pole approximation
assumption respectively,  $|\rho_I^0\rangle\equiv|1,0\rangle$ and
$|\omega_I\rangle\equiv|0,0\rangle$ are in the isospin basis, and
$\Pi_{\rho\omega}\equiv\langle \rho_I^0|W|\omega_I\rangle$. Note
that $\epsilon$ is not a real number, and hence the transfer
matrix from isospin basis to physical basis is not unitary. The
physical states are normalized in following way

\begin{equation}\label{norm}
\langle \tilde{a}|b\rangle =\delta_{ab}\;\;\;{\rm and}\;\;\;
\;{\bf I}=\sum_{a} |\tilde{a} \rangle \langle a|=\sum_a|{a}
\rangle \langle \tilde{a}|
\end{equation}
where $a,\;b=\rho^0_p$ or $\omega_p$. Because the transfer matrix
between the isospin bases and the physical one is not unitary in
the pole-MP,  the relations of eq.~(\ref{5}) can not be realized
in the formalism of unitary Lagrangian field theory. Since it has
been widely accepted that the effective Lagrangian field theories
inspired by QCD conceptions (e.g., VMD, large $N_c$ expansion,
non-linear realization of chiral symmetry, WCCWZ realization,
WZ-anomaly, hidden local symmetry and so on) is legitimate and
useful to study $\rho-$ and $\omega-$ physics\cite{eff}, that the
pole-MP approach contradicts to effective Lagrangian field theory
is a difficulty for pole-MP.

To show the differences between two approach  further, let us
consider $\rho^0,\omega\to\pi^0\gamma$ decays in the pole-MP.
Again, using $f_{\rho^0\pi^0\gamma},\;f_{\omega\pi^0\gamma}$ and
$f^{(0)}_{\rho^0\pi^0\gamma},\;f^{(0)}_{\omega\pi^0\gamma}$ to
denote the physical coupling constants and the isospin-basis
constants respectively, we have
\begin{eqnarray}\label{7}
f_{\rho^0\pi^0\gamma}|_{\rm pole-MP}
&=&f^{(0)}_{\rho^0\pi^0\gamma}(1
  +\frac{3\Pi_{\rho\omega}(m_\rho^2)}
   {m_\rho^2-m_\omega^2+
   i(m_\omega\Gamma_\omega-m_\rho\Gamma_\rho)}),
    \nonumber\\
f_{\omega\pi^0\gamma}|_{\rm pole-MP}
&=&f^{(0)}_{\omega\pi^0\gamma}(1
  +{1\over 3}\frac{\Pi_{\rho\omega}(m_\omega^2)}
   {m_\omega^2-m_\rho^2-i(m_\omega\Gamma_{\omega}-m_\rho\Gamma_\rho)}).
\end{eqnarray}
where $f^{(0)}_{\omega\pi^0\gamma}=3f^{(0)}_{\rho^0\pi^0\gamma}$
has been used. Comparing eq.~(\ref{1}) with eq.~(\ref{7}) and
noting $\Gamma_\omega<<\Gamma_\rho$, we can see
\begin{eqnarray}\label{10}
f_{\rho^0\pi^0\gamma}|_{\rm pole-MP} &\sim&
f^{(0)}_{\rho^0\pi^0\gamma} << f_{\rho^0\pi^0\gamma}|_{\rm eff-L}
\\
f_{\omega\pi^0\gamma}|_{\rm pole-MP} &\sim&
f^{(0)}_{\omega\pi^0\gamma}\sim f_{\omega\pi^0\gamma}|_{\rm
eff-L} \nonumber
\end{eqnarray}
Eq.~(\ref{10}) indicates there is no the hidden charge-asymmetry
enhancement effect for $\rho^0\rightarrow \pi^0\gamma $ mentioned
above in the pole-MP formalism. The result of pole-MP approach to
$\rho^0\rightarrow \pi^0\gamma $ is significantly different from
one of eff-L. Thus we learned that there may be something wrong.
The purpose of this letter is to re-examine the pole-Mixed
Propagator approach, and propose a new mixed propagator approach
based on the criterion of experimental measurement to meson
resonances.

Let's consider whether the pole-MP assumption on physical states
eq.~(\ref{4}) is consistent with the experimental criterion to
measure the physical meson resonance or not. We first consider the
single vector meson resonance channel case. The $D_{\mu\nu}$ of
eq.~(\ref{2}) is now the ordinary propagator of a vector meson.
The reaction amplitude for a process by the medium of this vector
meson resonance reads
\begin{equation}\label{11}
{\cal M}\sim J_1^\mu D_{\mu\nu}J_2^\nu=(J_1\cdot J_2){1\over
s-W(s)}
\end{equation}
where $J_1^\mu$ and $J_2^\nu$ represent some currents and $q_\mu
J_{1,\; 2}^\mu=0$. The reaction probability is\cite{Peskin}
\begin{equation}\label{12}
\sigma (s)\sim |{\cal M}|^2\propto |{1\over s-W(s)}|^2
\end{equation}
The meson resonance mass measured in the experiment is real and
is determined by the location of the maximum of $\sigma(s)-$peak
in real $s-$axis. Then the resonance mass is determined by the
following equation
\begin{equation}\label{13}
{\pa \over \pa s} |{ s-W(s)}|^2=0
\end{equation}
Consequently, the mass-square of the resonance, $M^2$, is
determined by the solution of above equation, i.e.,
\begin{equation}\label{14}
s=[ReW-{1\over 4}{\pa\over \pa s}(W^*W+WW^*)](1-{\pa\over \pa
s}ReW)^{-1}\equiv M^2(s).
\end{equation}

Now return to $\rho^0-\omega$ two channel case, $W(s)$ and hence
$M^2(s)$ are $2\times 2$ matrices. The mass determination
equation for physical resonance states should read
\begin{equation}\label{15}
det[s-M^2(s)]=0.
\end{equation}
The physical states then are the eigenvectors of the real
mass-matrix $M^2$. Obviously, the corresponding pole-MP
equation~(\ref{4}) is definitely different from above mass
determination equation derived from the experiment measurement
principle to physical resonance states. It indicates that the the
pole-MP is not consistent with the criterion of experimental
measurement to the physical resonances. Consequently, we should
use eq.~(\ref{15}) to define true physical $\rho-$ and
$\omega-$fields, and rule out the condition of eq.~(\ref{4}). For
convenience, we will call this method hereafter as
Experiment-criterion Mixed Propagator approach, or expt-MP
approach.

Now let us use expt-MP to fit experiments. Following refs\cite
{oconnell,renard} and using Breit-Wigner approximation, we have
\begin{eqnarray}
W=\left(
\begin{array}{lcr}
m_{\rho_I}^2-i\sqrt {s} \Gamma_{\rho_I} & \Pi_{\rho\omega}(s) \\
\Pi_{\rho \omega}(s) & m_{\omega_I}^2-i \sqrt{s} \Gamma_{\omega_I}
\end{array}
\right ) \nonumber
\end{eqnarray}
Considering Im$\Pi_{\rho \omega}$ is small and hence
ignorable\cite{Gard}, we get
 \begin{eqnarray}
 ReW=\left(
 \begin{array}{lcr}
m_{\rho_I}^2 & \Pi_{\rho\omega}(s)  \\
\Pi_{\rho \omega}(s) & m_{\omega_I}^2
\end{array} \right ) \nonumber
 \end{eqnarray}
 Then, in $\partial ReW/ \partial s$,  only
off-diagonal elements of ${\partial \Pi_{\rho \omega}/ \partial
s}$ is left. Generally, for a broad class of models
$\Pi_{\rho\omega}(s)$ at $(\rho-,\omega-)$resonance energy region
can be determined by taking VMD-type $\rho-\omega$ mixing
Lagrangian  ${\cal
L}_{\rho\omega}=f_{\rho\omega}\rho^{\mu\nu}\omega_{\mu\nu}$ (
$V^{\mu\nu}=\pa^\mu V^\nu-\pa^\nu V^\mu,\ V=\rho,\omega$). It
leads to $\Pi_{\rho\omega}(s)=f_{\rho\omega}s$ which satisfies
$\Pi(s=0)_{\rho\omega}=0$ required by generic consideration in
ref{\cite{O'Connell94}}. Thus, we have
\begin{equation} \label{17}
{\partial \over {\partial s}}\Pi_{\rho \omega}=f_{\rho\omega}
    ={\Pi_{\rho \omega} \over s}|_{s\sim m^2_{\rho}}\simeq 6.7\times 10^{-3} \ll 1
   \ \Longrightarrow\ 1-({\partial ReW}/{\partial s})\simeq 1.
\end{equation}
where $|\Pi_{\rho\omega}| \simeq 4000MeV^2 $\cite{O'Connell94} has
been used in the estimation. Furthermore, noting
$\Gamma_{\omega}/m_{\omega}\ll 1$, we have
\begin{eqnarray}
 M^2 = \left(
\begin{array}{cc}
m_{\rho_I}^2-{\partial(s \Gamma_{\rho}^2) \over \partial s} &
     \Pi_{\rho \omega}-{1\over2}(m_{\rho_I}^2+m_{\omega_I}^2)
     {\partial \over \partial s} \Pi_{\rho \omega} \\
\Pi_{\rho \omega}-{1\over2}(m_{\rho_I}^2+m_{\omega_I}^2){\partial
\over \partial s}
    \Pi_{\rho \omega} &
       m_{\omega_I}^2
\end{array}
\right)
\end{eqnarray}
where the off-diagonal elements of $M^2$ matrix represent the
$\rho-\omega$ mixing in the isospin basis. The physical $\rho$
and $\omega$ are eigenstates of $M^2$. Because $M^2$ matrix is
real and symmetric, distinguishing from the transformation from
isospin basis to physical basis in the pole-MP, the corresponding
transfer-matrix of expt-MP is unitary. $M^2$ matrix is
diagonalized by the unitary $2\times2$ matrix $C$:
\begin{eqnarray}
C M^2 C^\dag =\left(
\begin{array}{lcr}
m_{\rho}^2  & 0 \\
0 & m_{\omega}^2
\end{array}
\right )\nonumber
\end{eqnarray}
where
\begin{eqnarray}\label{18}
C=\left(
\begin{array}{lcr}
1& -\eta \\
\eta & 1
\end{array}
\right),\hspace{0.5in}
\eta = -{\Pi_{\rho \omega}(1-{1\over 2s}
    (m_{\rho}^2+m_{\omega}^2)) \over (m_{\omega}^2-m_{\rho}^2)}.
\end{eqnarray}
Consequently, the solutions of expt-MP's physical state condition
of eq.~(\ref{15}) are follows
\begin{eqnarray}\label{501}
|\rho_p^0\rangle&=&|\rho_I^0\rangle -\eta |\omega_I\rangle,
\;\;\langle {\rho}^0_p|=|\rho_p^0 \rangle^\dag
 \nonumber \\
|\omega_p\rangle&=&|\omega_I\rangle +\eta |\rho_I^0\rangle,
\;\;\langle \omega_p|=|\omega_p \rangle^\dag.
\end{eqnarray}
Under this transformation, we have
\begin{eqnarray}\label{19}
CWC^{\dag}& = & \left(
\begin{array}{lcr}
 z_{\rho}& T\\
T & z_{\omega}
\end{array}
\right )
\end{eqnarray}
where $z_{\rho}=m_{\rho}^2-im_{\rho}\Gamma_{\rho},z_{\omega}=
m_{\omega}^2-im_{\omega}\Gamma_{\omega}$, and $T=\Pi_{\rho \omega}
-\eta(z_{\omega}-z_{\rho})$ are all defined in the physical state
basis of expt-MP. The propagator function in the physical basis
$D^{P}$ reads
\begin{eqnarray}\label{191}
D^P(s) &=&
C(s-W)^{-1}C^{\dag}=(s-CWC^{\dag})^{-1} \nonumber \\
 & = & \left(
\begin{array}{lcr}
(s-z_{\rho})^{-1}& (s-z_{\rho})^{-1} T (s-z_{\omega})^{-1}\\
(s-z_{\omega})^{-1} T(s-z_{\rho})^{-1} & (s-z_{\omega})^{-1}
\end{array}
\right )\nonumber \\
&\equiv&\left(
\begin{array}{lcr}
D^P_{\rho\rho} &D^P_{\rho\rho}T D^P_{\omega\omega}\\
D^P_{\omega\omega} T D^P_{\rho\rho} & D^P_{\omega\omega}
\end{array}
\right ) \equiv\left(
\begin{array}{lcr}
D^P_{\rho\rho} &D^P_{\rho\omega}\\
D^P_{\omega\rho} & D^P_{\omega\omega}
\end{array}
\right ).
\end{eqnarray}

For $V-X$-vertex ($V=\rho,\;\omega$ and $X$ represents other
particles), $f^F_{VX}$
 denotes the corresponding form-factor, $f^P_{VX}$ and $f^0_{VX}$
 denote the coupling constants in the physical basis and in the
 isospin basis respectively.
\begin{figure}[hptb]
    \centerline{\psfig{figure=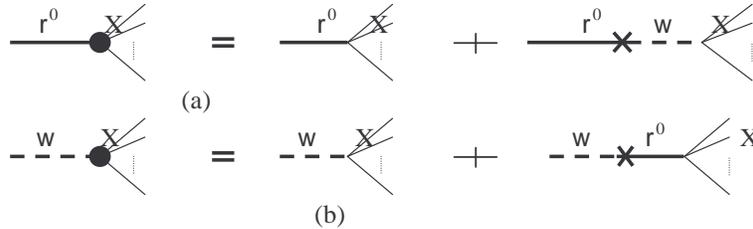,width=4in}}
 \centering
 \begin{minipage}{4.9in}
  \caption{The relation between the form factor of $V-X$ vertex and the corresponding
  coupling
  constants for the vertices, where $V=\rho$ in Fig.(a) and
  $V=\omega$ in Fig.(b).
  In the left-hand side, the black dots in the vertex represent the form factor.
  In the right-hand side, the coupling constants are at the vertices.
  The single thick (dash) lines denote for $\rho-$propagator
  $D^P_{\rho\rho}$
   ($\omega-$propagator $D^P_{\omega\omega}$), and the thick-cross-dash
   lines (or dash-cross-thick) line denote the mixed propagator
    $D^P_{\rho\omega}$
    (or $D^P_{\omega\rho}$). The thin lines are external lines of $X$-particles.}
 \end{minipage}
\end{figure}
 Since $D^P_{\rho\omega}=D^P_{\omega\rho}\not= 0$ in the
expt-MP, generally, the form factor is different from the
corresponding coupling  constant, i.e., $f^F_{VX}\not=f^P_{VX}$
(by contrast, $f^F_{VX}=f^P_{VX}$ in pole-MP approach due to
$D^P_{\rho\omega}=D^P_{\omega\rho}= 0$).  From Fig.1, we have
\begin{equation}\label{ffactor}
D^P_{VV}f^F_{VX}=D^P_{V\rho}f^P_{\rho X}+D^P_{V\omega}f^P_{\omega
 X},
\end{equation}
with
\begin{eqnarray}\label{602}
f^{P}_{\rho X}&=&f_{\rho X}^{(0)}-\eta
  f_{\omega X}^{(0)}\hspace{1in}
f^{P}_{\omega X}=f_{\omega X}^{(0)}+\eta
  f_{\rho\gamma}^{(0)}.
\end{eqnarray}

In terms of eqs.~(\ref{191}) and (\ref{602}) the time-like EM pion
form-factor is given, in the $\rho-\omega$ interference region, by
\begin{eqnarray}\label{603}
F_\pi (s)&=&1+[f^{P}_{\rho\gamma}D^P_{\rho\rho}f^{F}_{\rho\pi\pi}
+f^P_{\omega\gamma}D^P_{\omega\omega}f^F_{\omega\pi\pi}]\nonumber \\
&=& 1+{f^P_{\rho\gamma} f^P_{\rho\pi\pi}\over s-z_{\rho}}
    +{f^P_{\omega\gamma}f^P_{\omega\pi\pi}\over s-z_{\omega}}
    +{(f^P_{\rho\gamma}f^P_{\omega\pi\pi} + f^P_{\omega \gamma}f^P_{\rho\pi\pi})T
        \over (s-z_{\rho})(s-z_{\omega})}\nonumber \\
  &=&1+(f^P_{\rho\gamma} f^P_{\rho\pi\pi}+{(f^P_{\rho\gamma}f^P_{\omega\pi\pi}
    + f^P_{\omega \gamma}f^P_{\rho\pi\pi})T \over z_{\rho}-z_{\omega}})
   ({1 \over s-z_{\rho}}+\xi e^{i \phi}{1 \over s-z_{\omega}}),
\end{eqnarray}
with
\begin{eqnarray}\label{604}
\xi e^{i\phi}&=&[{1\over 3}\eta-{{{(\eta+{1\over 3})T}\over
{z_\rho -z_\omega}}}][1+{{(\eta+{1\over 3})T}\over {z_\rho
-z_\omega}}]^{-1},
\end{eqnarray}
where $f^{(0)}_{\rho\gamma}=3f^{(0)}_{\omega\gamma}$ and
$f^{(0)}_{\omega\pi\pi}=0$ have been used, and $\phi$ is Orsay
phase. Using $\Pi_{\rho\omega}\simeq -4000MeV^2$\cite{O'Connell94}
in eq.~(\ref{604}), we obtain that $\xi \simeq 0.012$ and $\phi$
is equal to about $100^o\rightarrow 101^o$ as $s$ varies from
$m_{\rho}^2$ to $m_{\omega}^2$. These predictions are in good
agreement with experimental data\cite{data}, and hence the
expt-MP approach is legitimate to describe the $\rho-\omega$
mixing effects in the pion EM-form factor.

\begin{figure}[hptb]
    \centerline{\psfig{figure=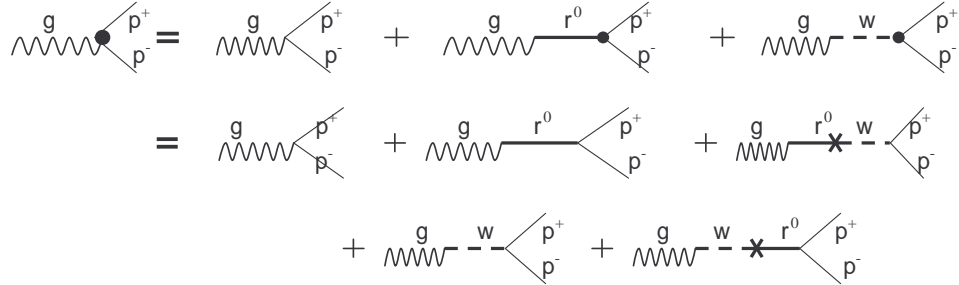,width=5in}}
 \centering
 \begin{minipage}{4.5in}
  \caption{The Electromagnetic(EM) pion form factor. Indications of the lines for
  the propagators are same as Fig.1. The black dots are form factors.}
 \end{minipage}
\end{figure}

Now, we study the anomalous-like $\rho^0 \rightarrow \pi^0
\gamma$ and $\omega \rightarrow \pi^0 \gamma$ decays in terms of
the expt-MP approach. Namely, taking $X=\pi^0\gamma$ in
eq.~(\ref{ffactor}), we have
\begin{eqnarray} \label {701f}
D^P_{\rho\rho}f^F_{\rho^0\pi^0\gamma}&=&D^P_{\rho\rho}
f_{\rho^0\pi^0\gamma}^{P} +D^P_{\rho\omega}f_{\omega \pi^0\gamma}^{P},\nonumber \\
D^P_{\omega\omega}f^F_{\omega\pi^0\gamma}&=&D^P_{\omega\omega}
f_{\omega\pi^0\gamma}^{P} +D^P_{\omega\rho}f_{\rho^0
\pi^0\gamma}^{P},
\end{eqnarray}
and hence
\begin{eqnarray} \label {701ff}
f^F_{\rho^0\pi^0\gamma} &=&f_{\rho^0\pi^0\gamma}^{(0)}
    -\eta f_{\omega \pi^0\gamma}^{(0)}
    +{T \over m_{\rho}^2-m_{\omega}^2
    +im_{\omega}\Gamma_{\omega}}f_{\omega \pi^0\gamma}^{(0)},\nonumber \\
f^F_{\omega\pi^0\gamma} &=&f_{\omega \pi^0\gamma}^{(0)}
    +\eta f_{\rho^0\pi^0\gamma}^{(0)}
    +{T \over m_{\omega}^2-m_{\rho}^2
    +im_{\rho}\Gamma_{\rho}}f_{\rho \pi^0\gamma}^{(0)},
\end{eqnarray}
where eqs.(~\ref{501}) and (~\ref{191}) have been used.
Considering $|\eta|\simeq 0.006<<1$ and $T\simeq
\Pi_{\rho\omega}$, we finally obtain the desired results in the
expt-MP formalism as follows
\begin{eqnarray} \label {701}
f_{\rho^0\pi^0\gamma}|_{\rm expt-MP}&\equiv&
f^F_{\rho^0\pi^0\gamma}\simeq f_{\rho^0\pi^0\gamma}^{(0)}
        +{\Pi_{\rho\omega} \over m_{\rho}^2-m_{\omega}^2
    +im_{\omega}\Gamma_{\omega}}f_{\omega \pi^0\gamma}^{(0)},\nonumber \\
f_{\omega\pi^0\gamma}|_{\rm expt-MP}&\equiv&
f^F_{\omega\pi^0\gamma}\simeq f_{\omega\pi^0\gamma}^{(0)}
        +{\Pi_{\rho\omega} \over m_{\omega}^2-m_{\rho}^2
    +im_{\rho}\Gamma_{\rho}}f_{\rho^0 \pi^0\gamma}^{(0)}.
\end{eqnarray}
Noting $f_{\omega \pi^0\gamma}^{(0)}=3f_{\rho^0
\pi^0\gamma}^{(0)}$, above equations are same as eq.(1).
Consequently, this result indicates that the charge-asymmetry
enhancement effect to $\rho^0\rightarrow \pi\gamma$ revealed in
ref.\cite{wjy} by using the effective Lagrangian theory has been
confirmed in expt-PM approach.

To conclude, the Mixed Propagator (MP) approach to $\rho-\omega$
mixing is investigated in this letter. It is found that under the
pole-approximation assumption the results of MP approach is not
compatible both with the effective Lagrangian theory and with the
experiment measurement criterion. This fact indicates the pole
approximation for determining the physical basis of $\rho$ and
$\omega$ is inadequate and ad hoc. To cure these diseases, we
propose a new MP approach in which the physical states of $\rho$
and $\omega$ are determined by the requirement of experimental
measurement to meson resonance. In terms of this new MP approach,
the EM pion form factor $F_\pi$ and form factors of $\rho^0
\rightarrow \pi^0 \gamma $ and of $\omega \rightarrow \pi^0
\gamma $ are derived. The results of $F_\pi$ are in good
agreement with data. The form factor of $\rho^0 \rightarrow \pi^0
\gamma $ exhibits a hidden charge-asymmetry enhancement effect
which agree with the prediction of the effective Lagrangian
theory, i.e., the conclusion of ref.\cite{wjy} has been confirmed.

We are pleased to acknowledge K.F.Liu for helpful discussion.

\end{document}